\documentclass[12pt]{iopart}

\RequirePackage{graphicx}
\usepackage{latexsym}
\expandafter\let\csname equation*\endcsname\relax
\usepackage{xcolor}
\expandafter\let\csname endequation*\endcsname\relax
\usepackage{amssymb}
\usepackage{amsmath} 
\usepackage{todonotes}
\begin{document}

\title{Fractional Schwarzschild-Tangherlini black hole with a fractal event horizon}

\author{S. Jalalzadeh$^{1,2}$\footnote{Author to whom any correspondence should be addressed}, H. Moradpour$^3$, G. R. Jafari$^{4,5}$, and  
P. V. Moniz$^{6,7}$}
\address{$^1$Departamento de F\'isica, Universidade Federal de Pernambuco, Recife, PE, 50670-901, Brazil}
\address{$^2$Department of Physics,  Dogus University, Dudullu-\"{U}mraniye, 34775 Istanbul, T\"{u}rkiye}
 \address{$^3$ Research Institute for Astronomy and Astrophysics of Maragha (RIAAM),
 University of Maragheh, P. O. Box 55136-553, Maragheh, Iran}
 \address{$^4$Physics Department, Shahid Beheshti University 1983969411, Tehran, Iran}
 \address{$^5$Center for Communications Technology, London Metropolitan University, London N7 8DB, UK}
\address{$^6$Centro de Matem\'atica e Aplica\c c\~oes (CMA-UBI), Covilh\~a, Portugal}
\address{$^7$Departamento de F\'{i}sica, Universidade da Beira Interior, 6200 Covilh\~a, Portugal}
\ead{shahram.jalalzadeh@ufpe.br, h.moradpour@riaam.ac.ir, gjafari@gmail.com, pmoniz@ubi.pt}

 \vspace{10pt}
\begin{indented}
\item[]March 2024
\end{indented}

\begin{abstract}
We demonstrate that the implementation of the fractional and non-local Wheeler--DeWitt (WDW) equation within the context of Schwarzschild geometry leads to the emergence of a Schwarzschild--Tangherlini black hole (BH), which is uniquely characterized by an event horizon that exhibits fractal properties and is defined by a non-integer dimension that lies in the continuum between the values of 1 and 2. Our calculations further reveal that this intriguing fractional BH may potentially possess a temperature that is substantially lower than that of a conventional BH, thereby suggesting a significant deviation from the expected thermodynamic properties of standard BHs. These remarkable characteristics, which are intrinsically linked to the non-integer dimensionality of the event horizon, likely arise from applying the Riesz fractional derivative as a sophisticated non-local operator, thus introducing fascinating dynamics into the theoretical framework of BH physics.
\end{abstract}

%%\submitto{\CQG}
%\maketitle

\section{Introduction} 
In 1974, about half a century ago, Hawking demonstrated (utilizing a seminal semiclassical framework) that BHs emit radiation due to quantum effects near their event horizons, leading to gradual mass loss and potential evaporation. More concretely, such radiation is of a black-body form, becoming the signature of the interplay between spacetime geometry and quantum fluctuations. Furthermore, entropy, $S$, was expressed as one-fourth of the event horizon area. It is, therefore, immensely tempting to further intertwine and explore Hawking radiation within a quantum gravity approach, trying new tools in the quest to understand the universe's most extreme conditions.

One such tool that recently gathered interest is fractional calculus \cite{Trivedi:2024inb, Bhoyar:2024gqy, Socorro:2024poa, 2024arXiv240517350S, 2023CSF7E, deOliveiraCosta:2023srx}: an extension of traditional calculus that allows for the modeling of complex systems and thus may be applied to describe phenomena in BH physics. In particular, by applying fractional derivatives with a non-integer order, we can suitably modify the WDW equation, which governs the quantum state, so as to gain access to discuss fractional dimensions and, most importantly, non-local effects. This suggests exploring the thermodynamics of a BH case study, which includes the mass spectrum, temperature, and entropy, using a fractional calculus lens. We can anticipate scale-dependent geometries, which can emerge subtly as quantum gravity effects on BH shadows and deflection of particle trajectories, for example, among several possibilities. 

In this letter, we utilized the fractional WDW equation to obtain the mass spectrum of a hole. This leads us to drive the entropy and the temperature of the fractional BH with a fractal event horizon, in which its fractal dimension depends on the l\'evy's parameter $\alpha$. Subsequently, we compute ($i$) the mass spectrum and the ground state mass in particular, ($ii$) the frequency of emitted or absorbed thermal radiation, ($iii$) the entropy as the surface area
of a ($D - 2$)-dimensional sphere, plus ($iv$) the effective $D$-dimensional gravitational constant,
($v$) the effective horizon radius, and ($vi$) the temperature of the fractional BH. Remarkably,
these quantities depend on the fractional parameter $0 < \alpha \leq  2$ that determines that the spacetime
dimension $D$ is expressed as $D \equiv  \frac{\alpha}{2} + 3$. Finally, we obtained the metric of the corresponding classical configuration as well as the Newtonian gravitational potential. The final section contains conclusions.

\section{Mass and entropy spectrum of a black hole}
The WDW equation for a Schwarzschild BH 
may be expressed as \cite{Jalalzadeh:2022rxx}
\begin{equation}
    \label{WDW1}
    -\frac{1}{2}\frac{d^2\psi}{dx^2}+\frac{1}{2}x^2\psi=\frac{2M^2}{m_\text{P}^2}\psi,
\end{equation}
where $\psi$ is the wave function of the BH, $M$ is its mass, and $m_\text{P}$ is the Planck mass in natural units, i.e., $\hbar=c=k_\text{B}=1$. Upon the use of (\ref{WDW1}),  the area of the event horizon, $A_n$, and the mass spectrum, $M_n$, are retrieved:
\begin{equation}\label{1-10}
%\begin{array}{cc}
A_n={8\pi l_\text{P}^2}(n+\frac{1}{2}),~~~
M_n=\frac{m_\text{P}}{\sqrt{2}}\sqrt{n+\frac{1}{2}},
%\end{array}
\end{equation}
where $n=0,1,2,...$, and $l_\text{P}=1/m_\text{P}$ is the Planck length (for details, see \cite{Jalalzadeh:2022rxx} and references therein).
The relations presented in (\ref{1-10}) yield the following results: On the one hand,  Hawking radiation occurs when a BH transitions from a higher state $n$ to a lower state $n'$, resulting in the emission of the difference in quanta. On the other hand, BHs do not undergo complete evaporation; instead, a remnant of Planck size remains following the evaporation process.

Assuming  Hawking radiation from a massive BH, i.e., \( M \gg m_\text{P} \) and \( n \gg 1 \), emitted during a transition from state \( n+1 \) to state \( n \), it is found \cite{Area1a, Xiang} that the BH dynamics proceeds between discrete mass eigenstates. Furthermore, the radiation is emitted in multiples of a fundamental frequency, approximately corresponding to the peak of the Hawking spectrum \cite{Coates:2019bun}.

Importantly, if we denote the frequency of emitted (or absorbed) radiation by $\omega$, then 
\begin{equation}\label{3-1}
    \omega=\gamma|\Delta M|\simeq\frac{\gamma}{2t_\text{P}\sqrt{2n}}\simeq\frac{\gamma m_\text{P}}{4Mt_\text{P}},~~~~n\gg1,
\end{equation}
where $\gamma$ is a dimensionless constant, $|\Delta M|=M_{n+1}-M_n$, and $t_\text{P}=1/m_\text{P}$ is the Planck time. This relation is supported by the correlation between the Bondi mass and outgoing radiation fluxes \cite{wald2010general}. Also, this is consistent with the classical BH oscillation frequencies, which scale as $1/M$. Therefore,  a BH will emit radiation at a characteristic temperature, $T\propto1/M$, corresponding to the Hawking temperature.

Let us now introduce a complementary perspective, which we will use later. In a physical system characterized by energy $E$ and vibrational frequency $\omega$, it is a simple task to demonstrate that the quantity
\begin{equation}\label{Invariant}
I=\int\frac{dE}{\omega},
\end{equation}
constitutes an adiabatic invariant \cite{Kunstatter:2002pj}. Through Bohr--Sommerfeld quantization, this invariant yields an equally spaced spectrum, expressed as \begin{equation}\label{I2}
I= 2\pi\left(n+\frac{1}{2}\right).
\end{equation}
In the case of our BH setting, if we utilize Eqs.  (\ref{1-10}),  (\ref{3-1}) and $E=M$, we find
\begin{equation}
    \label{H6}
    I=\int_{m_\text{P}}^M\frac{dM}{\omega}=\frac{2GM^2}{\gamma}=\frac{1}{\gamma}\left(n+\frac{1}{2}\right).
\end{equation}

Eqs. (\ref{I2}) and (\ref{H6}) are consistent if we choose $\gamma=1/(2\pi)$.
Therefore, regarding the definition of the Hawking--Bekenstein entropy, $S_\text{H-B}=A/(4G)$, and the second equality in (\ref{H6}), 
we can express the Hawking--Bekestein (H-B) entropy as the following adiabatic invariant
\begin{equation}
    \label{BHI}
    S_\text{H-B}=\int_{m_\text{P}}^M\frac{dM}{\omega}.
\end{equation}
%%%%%%%%%%%%%%%%%%%%%%%%%%%%%%%%%%%%%%%%%%%%%%%%%%%%%%%%%%%%%%%%%%%%%%%%%%%%%%%%%%%%%%
\section{Fractional black hole}
The fractional extension of the WDW equation (\ref{WDW1}) we will use follows from \cite{Junior:2023fwb, Jalalzadeh:2022uhl, Jalalzadeh:2020bqu} 
\begin{equation}
    \label{2+1}
    \frac{1}{2}\left(-\frac{d^2}{dx^2}\right)^\frac{\alpha}{2}\psi+\frac{1}{2}x^2\psi=\frac{2M^2}{m_\text{P}^2}\psi,~~~~0<\alpha\leq2,
\end{equation}
where $\alpha$ is the L\'evy's fractional parameter, and it is linked to the concept of L\'evy path \cite{Laskin:2002zz}. The fractional  Riesz derivative is defined in terms of the Fourier transformation \cite{Rie}
 \begin{equation}
 \left(-\frac{d^2}{dx^2}\right)^\frac{\alpha}{2} \psi(x)
=\displaystyle\frac{1}{\sqrt{2\pi}}
\int_{-\infty}^\infty dp e^{i{px}}
|p|^\alpha
\int_{-\infty}^\infty e^{-i{px'}}\psi(x')dx'.
\label{2ref8}
\end{equation}

The Riesz fractional derivative operates as a nonlocal operator unless \(\alpha = 2\). In fact, the fractional Laplacian in (\ref{2+1}) is affected by \(\psi(y)\) in proximity to \(x\) as well as by the values of \(\psi(y)\) across the entire minisuperspace. In the context of the fractional WDW equation, the behavior of the wave function in a specific region is influenced by the potential $x^2/2$ in that region and the potential throughout the entire space. Thus, the influence of quantum gravity on previously assumed solely classical domains may turn out to be judiciously substantial.

In fact, let us consider the momentum space, where the wave function may be expanded in a Fourier series.
%\begin{equation}
%    \label{Four}
%    \psi(x)=\frac{1}{\sqrt{2\pi}}\int_{-\infty}^\infty e^{-ipx}\phi(p)dp.
%\end{equation}
Then, the fractional WDW equation (\ref{2+1}) takes the following simple form \cite{2018Ph98e2127K}:
\begin{equation}
    \label{Four2}
    -\frac{1}{2}\frac{d^2\phi(p)}{dp^2}+\frac{1}{2}|p|^\alpha\phi(p)=\frac{2M^2}{m_\text{P}^2}\phi(p),
\end{equation}
where $\phi(p)$ is the Fourier transformation of $\psi(x)$.
Currently, there is no general analytical solution for equation (\ref{2+1}) or (\ref{Four2}), explicitly including a dependence on $\alpha$. However, in Ref. \cite{2018Ph98e2127K}, the author obtained the ground state and the first excited state by the variational method. On the other hand, we are interested in the semiclassical limit, where $M\gg m_\text{P}$. Therefore, we can 
instead employ the Bohr--Sommerfeld quantization condition to obtain the mass spectrum of the fractional BH, which from (\ref{1-10}) gives us 
\begin{equation}
    \label{SB}
    M_n=\frac{m_\text{P}}{{2}}\left(\frac{\pi}{B(\frac{1}{2},1+\frac{1}{\alpha})} \right)^\frac{\alpha}{\alpha+2}\left(n+\frac{1}{2} \right)^\frac{\alpha}{\alpha+2},
\end{equation}
where $B(a,b)$ is the Beta function, and $n$ is a large integer number \cite{Laskin:2002zz}.

Similarly to Eq. (\ref{3-1}), the frequency of emitted or absorbed thermal radiation now follows as
\begin{equation}
    \label{Rad1}
    \omega=\gamma|\Delta M|\simeq\frac{(D-3)\pi\gamma}{(D-2)2^\frac{D-2}{D-3}Bt_\text{P}}\left(\frac{m_\text{P}}{M}\right)^\frac{1}{D-3},
\end{equation}
where we defined
\begin{equation}
    \label{Dimension}
    D\equiv\frac{\alpha}{2}+3,~~~~~~3<D\leq 4.
\end{equation}
As a consequence of Hawking radiation emitted from the event horizon of a black hole, the fractional black hole will undergo a loss of mass until it attains its ground state mass.
As it is shown in Ref. \cite{2018Ph98e2127K}, by employing variational methods, we can retrieve the ground state of (\ref{Four2})
\begin{equation}
    \label{Ground}
    M_0=\sqrt{\frac{(D-3)\Gamma(\frac{D-3}{D-2})}{\Gamma(\frac{1}{D-1})}}\left(\frac{2\sqrt{(D-3)}}{D-2} \right)^\frac{2(3-D)}{D-2}m_\text{P}.
\end{equation}
\begin{figure}
    \centering
    \includegraphics[width=8cm]{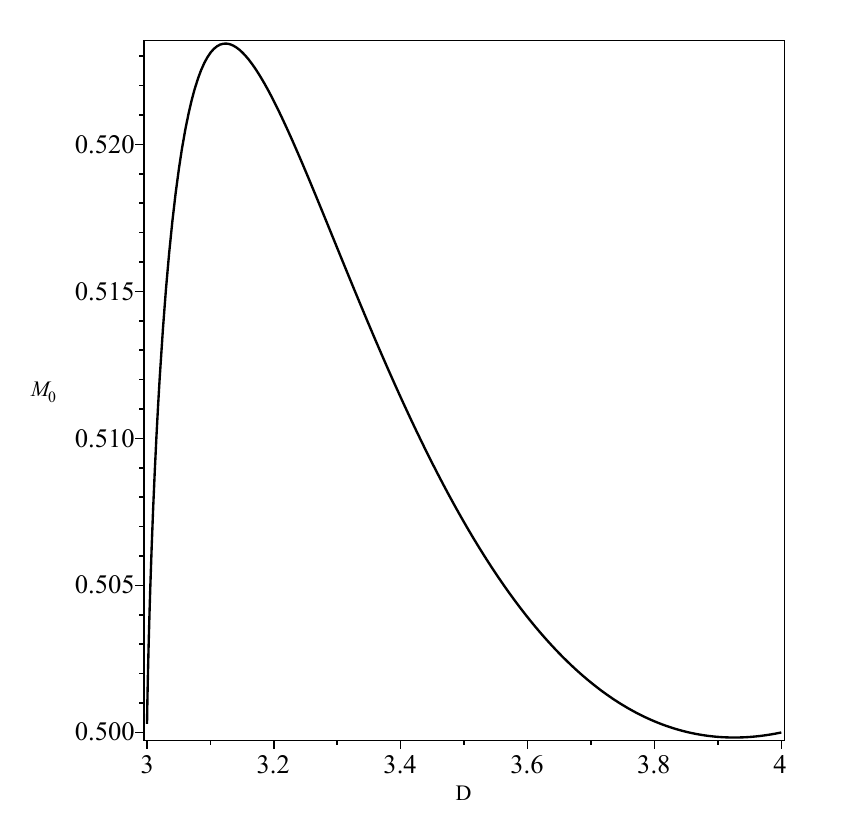}
    \caption{The ground state mass of a fractional black hole as a function of $D$ in the Planck units, i.e., $m_\text{P}=1$.}
    \label{Fig0}
\end{figure}
 As Fig. \ref{Fig0} shows, the ground state mass has a maximum at $D=3.1238$ with $M_0=0.5234m_\text{P}$, and a minimum at $D=3.9258$ with $M_0=0.4998m_\text{P}$. These are slightly different from ordinary ($D=4$) ground state mass given by Eq. (\ref{1-10}) with $M_0=0.5m_\text{P}$. 

Eq. (\ref{Rad1}) indicates that astrophysical BHs act as ``magnifying lenses,'' since they render the Planck-scale discretization of the horizon observable within the context of gravitational wave measurements. The emitted wave frequency is given by $f=\omega/(2\pi)\simeq f_\text{P}(m_\text{P}/M)^\frac{1}{D-3}$, with $f_\text{P}=2.95\times10^{42}$ Hz representing the Planck frequency. For $M=10M_\odot$, characteristic of the astrophysical BHs identified by LIGO-Virgo, the radiation frequencies for an ordinary BH $(D=4)$, for a fractional BH with the minimum ground state mass and maximum ground state mass are: $f(D=4)=128.1$ Hz, $f(D=3.9258)=10$ mHz, and $f(D=3.1238)=10^{-277}$ Hz, respectively. 

\section{Fractal event horizon} 
Employing the  results (\ref{SB})-(\ref{Dimension}) in the adiabatic invariant (\ref{BHI}), it allows  us to present  the entropy of the fractional BH as
\begin{equation}\label{Fblack1}
      S^\text{(Frac)}_{H-B}=\frac{\Omega_{(D-2)}R^{D-2}_S }{4\tilde G},
\end{equation}
where $\Omega_{(D-2)}=\frac{2\pi^\frac{D-1}{2}}{\Gamma(\frac{D-1}{2})}$ is the surface area of a $(D-2)$-dimensional unit sphere,
\begin{equation}\label{Fblack3}
    \tilde G=\frac{G^\frac{D-2}{2}(D-2)^{D-2}\Omega_{(D-2)}B(\frac{1}{2},
    \frac{2D-5}{2(D-3)})^{D-3}}{8\pi^{D-2}},
\end{equation}
is the effective $D$-dimensional gravitational constant, and
\begin{equation}\label{Fblack4}
    R_S=\left(\frac{16\pi\tilde GM}{(D-2)\Omega_{(D-2)}} \right)^\frac{1}{D-3},
\end{equation}
is the effective horizon (Schwarzschild--Tangherlin) radius.
Furthermore, comparing the differential of this relation with the first law of thermodynamics for BHs, \(dM=T_\text{H}dS\), provides the temperature of the fractional BH:
\begin{equation}
    \label{FracTemp}
    T^\text{(Frac)}_H=\frac{D-3}{4\pi R_S}.
\end{equation}
This relation indicates that the temperature of a fractional BH is highly dependent on the fractional parameter and decreases rapidly as \(D\) decreases. For instance, for a BH with mass $10M_\odot$, when \(D = 4\), the temperature \(T\) is approximately \(\mathcal{O}(10^{-9})\) K, for \(D = 3.9258\), \(T\) is around \(\mathcal{O}(10^{-12})\) K, and for \(D = 3.1238\), \(T\) drops to approximately \(\mathcal{O}(10^{-287})\) K. Generally, $0<T\leq\mathcal O(10^{-9})$ K, for $3<D\leq 4$. This suggests that the lifetimes of fractional astrophysical BHs could be significantly longer than those of ordinary BHs. Also, the heat capacity of the fractional BH is the same as the Schwarzschild--Tangherlin BH, given by $C=\frac{1}{M}\frac{dM}{dT^\text{(Frac)}_H}=\frac{3-D}{T^\text{(Frac)}_H}$. Regarding the constraint interval on the value of $D$ in (\ref{Dimension}), the heat capacity of the fractional BH is always negative, which shows that the BH is unstable.

Eqs. (\ref{Fblack1})-(\ref{Fblack4}) resemble the entropy of a $D$-dimensional  Schwarzschild--Tangherlini BH \cite{Tangherlini:1963bw, Kunstatter:2002pj}, in which its event horizon ``surface'' is a fractal with a dimension $D-2=\frac{\alpha}{2}+1$ and, pertinently, $D$ is related to the L\'evi's fractional parameter by Eq. (\ref{Dimension}). To discuss the impact of fractionality on the physics of the BH event horizon, we will revisit the original Schwarzschild BH as described by Eqs. (\ref{1-10}). An approach that assists in 
elucidating a meaning for the entropy of a BH is detailed in the article \cite{Khosravi:2010wq} and references therein.

\begin{figure}
  \centering
  \includegraphics[width=8cm]{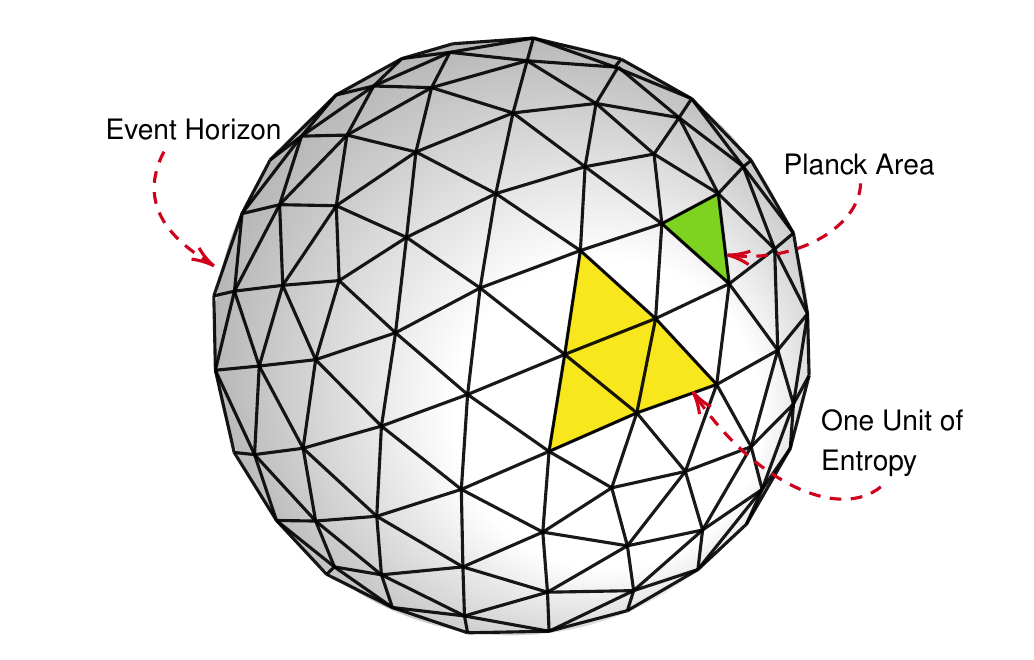}
  \caption{The figure schematically represents the standard interpretation of the entropy-area relationship for a black hole.}
   \label{Fig1}
\end{figure}

Consider a sphere with a radius \( R_S \) whose surface area is triangulated using Planck areas \( l_\text{P}^2 \) as illustrated in Fig. \ref{Fig1}.
%This figure illustrates
Then, one unit of information corresponds to four times the Planck area, whereas the Bekenstein--Hawking entropy of BHs suggests that information units constitute the total surface area of the event horizon.

\begin{figure}
    \centering
    \includegraphics[width=5cm]{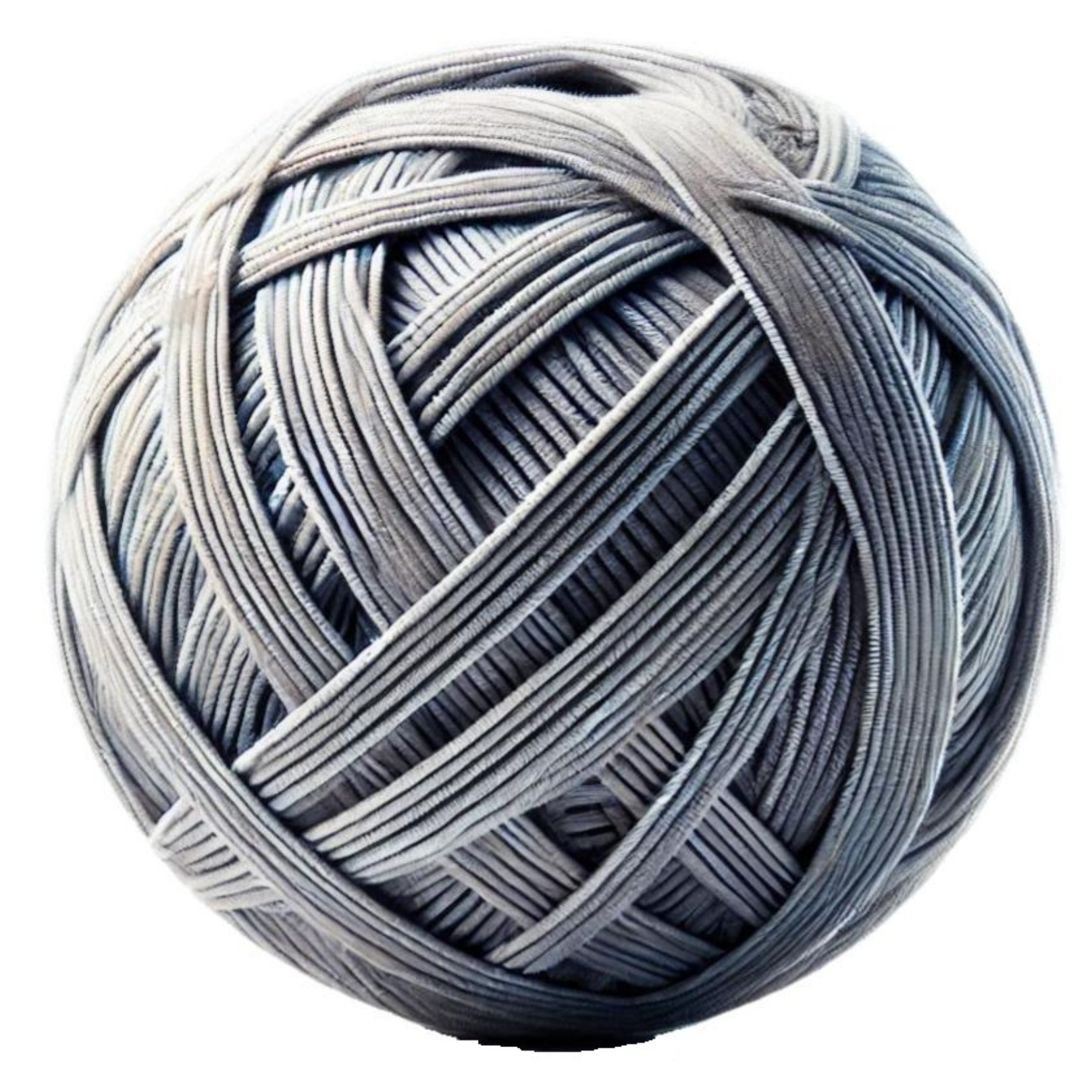}
    \caption{A cartoon schematically representing the entropy-area of a fractional black hole.}
    \label{Fig2}
\end{figure}

In contrast, the dimension of the ``surface'' area of the fractional horizon is a non-integer, constrained between the ranges $1<D-2\leq 2$. Consequently, the previous depiction is no longer applicable; instead of a triangulated surface, we may now imagine our perspective by means of a spherically shaped entanglement of yarn formed by concatenating Planck lengths, as seen in Fig. \ref{Fig2}, or we can consider it as a Swiss cheese with high porosity. As a result, the fractional dimension suggests that there are more hidden secrets than are described by integer-dimensional models.
Note that in the new fractional picture of the BH, $\alpha=2$, we will depict a triangled meshed area in place of the triangulated surface area at the location of the event horizon. 

%Another effect of the fractal nature of the horizon surface is the effect of non-local information from the horizon. In other words, information from different Planck regions would reach any point in space. Entropy is a measure of the disorder or randomness of a system. If the number of information units increases on the horizon, it would mean that there is more disorder or randomness in the system. This could impact the entropy of the horizon as the system becomes more chaotic and unpredictable.

The surface area of $(D-2)$-dimensional fractal unit sphere defined by (\ref{Fblack1}) can be explained by $\Omega_{(D-2)}=\int d\mu_H$, where $d\mu_H$ denotes an appropriate Hausdorff measure over the space \cite{tarasov2011fractional} defined by
\begin{equation}
    d\mu_H=\frac{\pi^\frac{D-3}{2}}{\Gamma(\frac{D-3}{2})}|\cos\theta|^{D-4}\sin\theta d\theta d\varphi,
\end{equation}
in the spherical coordinates $(\theta,\varphi)$. Also, the event horizon's line element is given by \cite{Varieschi:2021rzk}
\begin{equation}
    \label{Fline}
    dl_H^2=\Big(\cos^2\theta+\\\left(\frac{\pi^\frac{D-3}{2}}{\Gamma(\frac{D-3}{2})}\right)^2|\cos\theta|^{2(D-4)}\sin^2\theta \Big)d\theta^2+\sin^2\theta d\varphi^2.
\end{equation}

It is necessary to establish the relevant spacetime metric to examine the acquired BH more closely. To achieve this, given that the temperature is known~(\ref{FracTemp}), we utilize its relationship with the surface gravity in a static, spherically symmetric manifold to determine the metric components. For this aim, let us consider the static spherically symmetric 
metric\footnote{{Please notice that our choice of metric signature is $(-,+,+,+)$.}}
\begin{equation}\label{h2}
ds^2=-f(r)dt^2+\frac{dr^2}{f(r)}+r^2dl_H^2,
\end{equation}
whose surface gravity, at a hypothetical horizon located at $R_S$, is calculated using
$\kappa=\frac{1}{2}|f^{\prime}(r)_{r=R_{S}}|$,
 where a prime denotes radial derivative. Since $\kappa=2\pi T^\text{(Frac)}_H$, we find
\begin{equation}\label{h4}
f(r)=1-\frac{16\pi\tilde GM}{(D-2)\Omega_{(D-2)}r^{D-3}}.
\end{equation}
Eqs. (\ref{Fline}), (\ref{h2}) and (\ref{h4}) convey a fractal extension of Schwarzschild--Tangherlini BH with a non-integer dimension given by Eq. (\ref{Dimension}). 

The form of the time-time component of the metric leads us to the Newtonian gravitational potential. Regarding the fact that $g_{00}=-1-2V(r)$, where $V(r)$ is the gravitational potential, one easily finds
\begin{equation}
    \label{Pot}
    V(r)=-\frac{8\pi\tilde GM}{(D-2)\Omega_{(D-2)}r^{D-3}}.
\end{equation}

Several studies have examined similar gravitational potentials that vary in the power of $r$ \cite{Giusti:2020rul, Varieschi:2020hvp, Moradpour:2024uqa}. The differences arise from the methods used to incorporate fractional considerations. Measurements show that on conventional scales, the Newtonian potential, which is inversely proportional to the distance, functions effectively, i.e., $D=4$ in Eq. (\ref{Pot}). Gravity has only been precisely measured at scales within the 1 cm range. It is uncertain if gravity remains unchanged at sizes lower than 1 cm. Conversely, interstellar measurements indicate that $2.9<D\leq 4$ \cite{Moradpour:2024uqa}. Furthermore, on a galactic scale, to account for rotation curves without presupposing the presence of dark matter $D \approx 3$ \cite{Giusti:2020rul, Varieschi:2020hvp}. It may be concluded that accurately characterizing the transition among these three asymptotic regimes (in other words, the scale-dependent behavior of gravitational potential (\ref{Pot})) necessitates treating the fractional WDW equation (\ref{2+1}) as a variable-order fractional differential equation, with $\alpha=\alpha(r)$.

\section{Conclusion}
Fractional calculus has demonstrated its use in analyzing many physical systems, including cosmology \cite{FQ2, FQ3, Jalalzadeh:2024qej, Junior:2023fwb} and quantum gravity \cite{FracBH, Jalalzadeh:2022uhl, Rasouli:2022bug, bidlan2025reconstructing, 
doi:10.1142/8540, 
fractalfract9060349, Calcagni:2021aap}. In this letter, we show that a fractional extension of the WDW equation provides a new understanding of the Schwarzschild BH. We showed that while the geometrical properties and the thermodynamics of the resulting BH exhibit similar quantities of a Schwarzschild--Tangherlin BH, its horizon is a fractal with a dimension between $1<D-2\leq2$. Our calculations show that the temperature of a fractional BH could be extremely lower than the ordinary BH, which means these BHs may be almost eternal. Since the fractional WDW equation is a non-local equation, the influence of fractionality is not negligible even at the classical level. As a result, the obtained line element and the corresponding Newtonian gravitational potential of a pointlike particle are affected by the fractional parameter.

Our findings present a significant opportunity to rigorously reassess classical tests of general relativity within the solar system. Additionally, they highlight the intricacies associated with the BH shadow and the gravitational wave echoes by BHs.
 It is essential to consider the spin of the BH when addressing any claims regarding astrophysical BHs, as actual BHs exhibit rotation. Consequently, in our next research, we will expand our investigation to encompass the Kerr BH.

\ack
S.J. acknowledges financial support from the National Council for Scientific
and Technological Development -- CNPq, Grant no. 308131/2022-3. {PVM (ORCID 0000-0001-7170-8952) acknowledges the FCT grant UID-B-MAT/00212/2020 at CMA-UBI plus
the COST Actions CA23130 (Bridging high and low energies in search of
quantum gravity (BridgeQG)) and CA23115 (Relativistic Quantum Information (RQI)).}
\vspace{.3cm}
\section*{References}
\bibliographystyle{iopart-num}       
\bibliography{Frac}   % name your BibTeX data base

%%%%%%%%%%%%%%%%%%%%%%%%%%%%%%%%%%%%%%%%%%%%

\end{document}